\documentclass[conference]{IEEEtran}
\IEEEoverridecommandlockouts
\usepackage{cite}
\usepackage{amsmath,amssymb,amsfonts}
\usepackage{algorithmic}
\usepackage{graphicx}
\usepackage{textcomp}
\usepackage{color,soul} 

\usepackage{verbatim}

\usepackage{xspace}
    
  \usepackage{listings}%

\usepackage{balance}

\usepackage[usenames,dvipsnames]{xcolor}
\usepackage{colortbl}
\definecolor{TableRowGray}{gray}{0.93}
\newcommand{\vect}[1]{\ensuremath{\boldsymbol{#1}}}

\newcommand{\vp}{\vect{p}}
\newcommand{\vpi}{\vect{\pi}}

\newcommand{\calB}[0]{\ensuremath{\mathcal{B}}}
\newcommand{\calF}[0]{\ensuremath{\mathcal{F}}}
\newcommand{\calM}[0]{\ensuremath{\mathcal{M}}}
\newcommand{\calS}[0]{\ensuremath{\mathcal{S}}}

\newtheorem{example}{Example}    
\definecolor{keywords}{HTML}{8A4A0B}

\definecolor{background}{HTML}{EEEEEE}

\definecolor{comments}{HTML}{868686}
\lstdefinelanguage{riaps}{
 morekeywords={ package,  struct, sequence,req,pub,sub, component,long,float, timer, char,double, client,string, server, requires, nodeLabel, nodeTemplate, artifact, device, located, at, system, nodes,node, key,components,import,sameactor,library collocate,distribute, =>,provides, memory, MB, storage,app, deploy, objective, message,'{','}'},
 keywordstyle=\color{keywords},
    basicstyle=\scriptsize\ttfamily,
 morecomment=[l]{//}, 
 morecomment=[s]{/*}{*/}, 
 morestring=[b]",
    basicstyle=\scriptsize\ttfamily,%
 commentstyle=\color{comments}\ttfamily,
numbers=right,
    numberstyle=\scriptsize,
    stepnumber=1,
    numbersep=10pt,
breaklines=true,
    frame=tb,
 tabsize=4}

\usepackage{tikz}
\usetikzlibrary{shadows}
\usepackage{pgfplots}
\pgfplotsset{compat=1.10}

\definecolor{blueLine}{RGB}{57,106,177}
\definecolor{blueFill}{RGB}{114,147,203}
\definecolor{redLine}{RGB}{204,37,41}
\definecolor{greenline}{RGB}{0,250,0}
\definecolor{blackLine}{RGB}{0,0,0}
\definecolor{goldLine}{RGB}{160,82,45}

\usepackage[textsize=tiny, textwidth=1.455cm, disable]{todonotes}

\newcommand{\Aron}[1]{\todo[color=red!20, linecolor=black!50]{\textbf{Aron:} #1}}

\definecolor{verylightgray}{rgb}{.97,.97,.97}

\lstdefinelanguage{Solidity}{
	keywords=[1]{anonymous, assembly, assert, balance, break, call, callcode, case, catch, class, constant, continue, contract, debugger, default, delegatecall, delete, do, else, emit, event, export, external, false, finally, for, function, gas, if, implements, import, in, indexed, instanceof, interface, internal, is, length, library, log0, log1, log2, log3, log4, memory, modifier, new, payable, pragma, private, protected, public, pure, push, require, return, returns, revert, selfdestruct, send, storage, struct, suicide, super, switch, then, this, throw, transfer, true, try, typeof, using, value, view, while, with, addmod, ecrecover, keccak256, mulmod, ripemd160, sha256, sha3}, % generic keywords including crypto operations
	keywordstyle=[1]\color{blue}\bfseries,
	keywords=[2]{address, bool, byte, bytes, bytes1, bytes2, bytes3, bytes4, bytes5, bytes6, bytes7, bytes8, bytes9, bytes10, bytes11, bytes12, bytes13, bytes14, bytes15, bytes16, bytes17, bytes18, bytes19, bytes20, bytes21, bytes22, bytes23, bytes24, bytes25, bytes26, bytes27, bytes28, bytes29, bytes30, bytes31, bytes32, enum, int, int8, int16, int24, int32, int40, int48, int56, int64, int72, int80, int88, int96, int104, int112, int120, int128, int136, int144, int152, int160, int168, int176, int184, int192, int200, int208, int216, int224, int232, int240, int248, int256, mapping, string, uint, uint8, uint16, uint24, uint32, uint40, uint48, uint56, uint64, uint72, uint80, uint88, uint96, uint104, uint112, uint120, uint128, uint136, uint144, uint152, uint160, uint168, uint176, uint184, uint192, uint200, uint208, uint216, uint224, uint232, uint240, uint248, uint256, var, void, ether, finney, szabo, wei, days, hours, minutes, seconds, weeks, years},	% types; money and time units
	keywordstyle=[2]\color{teal}\bfseries,
	keywords=[3]{block, blockhash, coinbase, difficulty, gaslimit, number, timestamp, msg, data, gas, sender, sig, value, now, tx, gasprice, origin},	% environment variables
	keywordstyle=[3]\color{violet}\bfseries,
	identifierstyle=\color{black},
	sensitive=false,
	comment=[l]{//},
	morecomment=[s]{/*}{*/},
	commentstyle=\color{gray}\ttfamily,
	stringstyle=\color{red}\ttfamily,
	morestring=[b]',
	morestring=[b]"
}

\begin{document}

\setlength{\marginparwidth}{1.4cm}

\title{TRANSAX: A Blockchain-based Decentralized Forward-Trading Energy Exchange for\\Transactive Microgrids
\thanks{The information, data, or work presented herein was funded in part by Siemens, CT and  Advanced Research Projects Agency-Energy (ARPA-E), U.S. Department of Energy, under Award Number DE-AR0000666.}}

\newcommand{\platform}[0]{TRANSAX\xspace}

\author{\IEEEauthorblockN{Aron Laszka}
\IEEEauthorblockA{\textit{Computer Science} \\
\textit{University of Houston}\\
Houston, TX}
\and
\IEEEauthorblockN{Scott Eisele, Abhishek Dubey, Gabor Karsai}
\IEEEauthorblockA{\textit{Institute for Software Integrated Systems} \\
\textit{Vanderbilt University}\\
Nashville, TN}
\and 
\IEEEauthorblockN{Karla Kvaternik}
\IEEEauthorblockA{\textit{Siemens Corporate Technology}\\
Princeton, NJ}
}
\pagenumbering{arabic}
\pagestyle{plain}
\maketitle

%\vspace{-8em}

\begin{center}
Published in the proceedings of the 24th IEEE International Conference on Parallel and Distributed Systems\\(IEEE ICPADS 2018).
\end{center}

\begin{abstract}
Power grids are undergoing major changes due to rapid growth in renewable energy and improvements in battery technology. Prompted by the increasing complexity of power systems, decentralized IoT solutions are emerging, which arrange local communities into transactive microgrids. The core functionality of these solutions is to provide mechanisms for matching producers with consumers while ensuring system safety. However, there are multiple challenges that these solutions still face: privacy, trust, and resilience. The privacy challenge arises because the time series of production and consumption data for each participant is sensitive  and may be used to infer personal information. Trust is an issue because a producer or consumer can renege on the promised energy transfer. Providing resilience is challenging due to the possibility of failures in the  infrastructure that is required to support these market based solutions. In this paper, we develop a rigorous solution for transactive microgrids that addresses all three challenges by providing an innovative combination of MILP solvers, smart contracts, and publish-subscribe middleware within a framework of a novel distributed application platform, called Resilient Information Architecture Platform for Smart Grid. Towards this purpose, we describe the key architectural concepts, including fault tolerance, and show the trade-off between market efficiency and resource requirements. 
%This paper describes a framework that address the problem of implementing transactive energy mechanisms in a distributed setting, providing privacy, resilience and trust. The key aspect of the framework 
%Specifically, we design and implement an automated auction and matching system that ensures safety (i.e. satisfaction of line capacity constraints), preserves privacy, and promotes local trade and market efficiency for IoT-based transactive energy systems
%This paper describes a state of the art transactive energy mechanism implemented over a decentralized middleware architecture, specially designed for smart grid. In the paper, we describe the key architectural concepts, which includes a distributed ledger based solution. We describe the fault-tolerance architecture and show the tradeoff between efficiency of the solutions vs the resource required. At runtime, the decentralized middleware solution can reconfigure the solver dynamically trading off operational goal against the efficiency goal.
\end{abstract}


\section{Introduction}
\label{sec:intro}
%\Aron{Abhishek, please check the acknowledgement footnote at the bottom of this page!}
Power grids are undergoing major changes due to the rapid adoption of renewable energy resources, such as wind and solar power \cite{EIA2014,5430489}. For example, 4,143 megawatts of solar panels were installed in the third quarter of 2016 \cite{seia}. This capacity is predicted to grow from 4\% of the total global energy production in 2015 to 29\% in 2040 \cite{Randal}. Simultaneously, battery technology costs per kWh have been dropping significantly \cite{stock2015powerful}, reaching grid parity \cite{bronski2015economics}. These trends are enabling a decentralized vision for the future of power-grid operations in which local peer-to-peer energy trading within microgrids can be used to reduce the load on distribution system operators (DSO), leading to the development of Transactive Energy Systems (TES) \cite{kok2016society,rahimi2012transactive, cox2013structured,melton2013gridwise}. Such mechanisms can improve system reliability and efficiency by integrating inverter-based renewable resources into the grid and by supplying power to the local loads when the main grid is interrupted.  

% Discuss that the energy exchange is a distributed cyber-physical systems problem and must be managed within a robust framework.
To accomplish the goal of transactive energy, individual prosumers\footnote{A prosumer is a home that can not only consume, but also produce surplus energy. Homes without production will be simply called consumers.} need to engage in interactions, negotiate with each other, enter agreements, and make proactive run-time decisions---individually and collectively---while responding to changing demands and environmental conditions. In theory, these interactions could happen in a centralized manner by communicating relevant variables to a central controller, which would compute and broadcast the ``optimal'' control settings back to each individual prosumer. However, this system would not scale well \Aron{To be fair, the blockchain based solution that we describe in this paper may scale even worse...} as the number of coordinating parties increases. It would also adversely affect resilience because of  increased risks for data corruption and loss during transmission. Further, the centralized controller would constitute a single point of failure. On the other hand, distributed optimization solutions might suffer from the same scalability challenges, and the ``distribution'' of the optimization problem often requires the over-simplification of objective functions, which would result in losing the guarantees of a globally ``optimal'' solution. 
In light of this, novel ``decentralized'' solutions are needed, in which individual prosumers operate with autonomous  controllers that can trade on their behalf in a market, which is itself decentralized. However, creating such decentralized solutions is challenging due to a number of problems.

The first problem is ensuring the physical stability and safety of the grid apparatus, which requires dynamically balancing supply and demand without violating line capacity constraints. The second one is a distributed systems problem, which requires ensuring that this peer-to-peer market operates in a trustworthy manner even if some of the nodes are malicious, compromised, or faulty. The third problem is related to privacy. 
While non-transactive smart metering systems require sharing prosumer information only with the DSO, transactive systems need to disseminate information among the participants to enable finding trade partners.
The dissemination of trading information threatens the privacy of prosumers since it  may expose their private information to anyone in the same microgrid. 
Further,
data collected from energy transactions is expected to be more fine-grained than data collected by currently deployed smart meters \cite{Privacy2017}, and it may be used to infer personal information about the market participants. For example, a participant's presence or absence at their residence might be inferable from their energy future offers (e.g., if a prosumer posts an energy selling offer, the residents are less likely to be at home). The fourth problem is resilience. Failures in distributed computing systems are a fact, and hence the transactive system must be able to tolerate failures by either mitigating faults or adapting the system to a different configuration.

\textbf{Contributions:} In this paper, we describe the design and implementation of \platform, a transactive decentralized platform built over a distributed middleware, called Resilient Information Architecture Platform for Smart Grid (RIAPS) \cite{eisele2017riaps,Scott2017ICCPS}. RIAPS isolates the hardware details from the algorithms and
provides essential mechanisms for resource management, fault
tolerance, and security. An integrated distributed ledger and smart contracts provide us with the mechanisms to provide consensus and trust. This is in line with the recent trends in the research community and industry focused on transactive energy markets \cite{Lo3Patent,PowerLedger}. Although disintermediation of trust is widely regarded as the primary feature of blockchain-based transaction systems \cite{SpectrumBC}, their use in TES is appealing also because they elegantly integrate the ability to immutably record the ownership and transfer of assets, with essential distributed computing services, such as Byzantine fault-tolerant consensus on the ledger state as well as event chronology. The ability to establish consensus on state and timing is important in the context of TES since these systems are envisioned to involve the participation of self-interested parties, interacting with one another via a distributed computing platform that executes  transaction management. We provide privacy by using a mixing service, which prevents tracing  assets being traded back to their owner, as described in our prior work \cite{Laszka17}\footnote{Note that since the mixing implementation was discussed in \cite{Laszka17}, in this paper we focus on other key contributions, including resilience.}. However, unlike \cite{Laszka17}, we consider an automated matching system that maximizes the amount of energy traded within the local market, while satisfying safety constraints. Finally, we describe and evaluate an extension to the RIAPS framework that implements distributed fault detection and mitigation mechanisms. These mechanisms are critical for resilient operation of \platform.

%we describe the resilience features of \platform and show how it can recover from failures.

\textbf{Outline}: The outline of this paper is as follows. We explain the problem of transactive energy systems using an example in Section \ref{sec:problem}. Then, we contextualize our contributions in \platform using related research in Section \ref{sec:related}. We describe \platform in Section \ref{sec:solution}, which is followed by an evaluation using a case study in Section \ref{sec:casetudy}. Finally, we conclude with discussions in Section \ref{sec:conclusion}.

\section{Transactive Energy Problem}
\label{sec:problem}
Consider a microgrid with a set of feeders arranged in a radial topology.\footnote{The methods developed in this paper are extensible to more general tree topologies involving branching. We work with a radial topology to simplify our notation.} 
A feeder\footnote{A feeder element in electrical distribution is a  power line transferring power from a distribution substation to distribution transformers or from distribution transformers to the end homes.} has a fixed set of nodes, each representing a residential load or a combination of load and distributed energy resources (DERs), such as rooftop solar and batteries. Each node is associated with a participant in the local peer-to-peer energy trading market. There is a distribution system operator (DSO), that also participates in the market and may thus use the market to incentivize timed energy production within the microgrid to aid in grid stabilization and promotion of related ancillary services \cite{7462854}. In addition, the DSO supplies residual demand not met through the local market.  The participants settle trades in advance, which allows them to schedule their transfer of power into the local distribution system. Alternatively, a mechanism can be responsible for matching the producers and consumers. There are typically three phases in these operations: discovery of compatible offers, matching of buying offers to selling offers (which may have been submitted either by each prosumer individually or by automated matching mechanisms). Once the matching is done, the energy transactions and financial transactions are then handled at a later time.

\begin{example}
\label{example:1}
Consider a community with two prosumers ($P_1$, $P_2$) and one consumer ($C_1$) on a single feeder.  To make the problem of matching energy offers tractable, lets assume that the offers are made and matched for discrete time intervals. These intervals quantize the whole day, and their length can be a parameter of the problem setup. For the sake of example, lets assume that each day is divided into 15 minute intervals. Lets assume that $P_1$ has the ability to transfer $10$ kW into the feeder during interval 48, which translates to 12:00pm--12:15pm. Assume similarly that $P_2$ can also provide  $30$ kW to the feeder in interval 48, but it has battery storage. Since $P_2$ has battery---unlike $P_1$, who must either transfer the energy or send the energy into the group---$P_2$ can delay the transfer until a future interval, e.g., interval 49.  Now suppose that $C_1$ needs to consume $30$ kW in interval 48 and $10$ kW in interval 49. All the prosumers and consumer must provide these requirements to the market mechanism, which will then  provide a matching solution. A possible solution would be to provide all $30$ kW to $C_1$ from $P_2$ in interval 48. However, that will lead to the waste of energy provided by $P_1$. Thus, a better solution will be to consume $10$ kW from $P_1$ in interval 48 and $20$ kW from $P_2$ in interval 48. Then, transfer $10$ kW from $P_2$ in interval 49, which is more efficient then the first matching as it allows more energy (summed across the intervals) to be transferred. Note that the second solution requires the market to consider future intervals while solving the problem, which increases the size of the optimization problem. Further, it should be noted that if the information about $C_1$'s offer is made public, then one can estimate that $C_1$ was doing heavy machine work during interval $48$ and there was substantially lower activity during interval $49$. 

%However, the actual result will depend upon if the market is myopic or not i.e. how many intervals it considers from future when solving the matching problem. 

%Even in this simple case, there are a number of possible solutions for interval 48. Ask $P_1$ to transfer all $10$ KW in interval 48 and $P_2$ provide the remaining 5 or let $P_2$ provide all the  15 KW for interval 48. It is clear that the second solution is inefficient because in that case $C_1$ will not find any match for interval 49 as $P_2$ will have exhausted all its capacity and $P_1$ cannot store energy beyond interval 48. This problem becomes even more challenging if the market is myopic i.e. it only performs matching one interval at a time. To generalize it is clear that the market might have an opportunity to be more efficient if it considers a large number of future intervals. However, the challenge is that a large number of future intervals will increase the size the matching problem, which as we will show later can be encoded as a Mixed Integer Linear Program (energy consumed in an interval is not discrete).
\end{example}

Based on this example, it is clear that there are five basic requirement that must be met by any solution.
\begin{itemize}
    \item The first requirement is the existence of an appropriate communication and messaging architecture. The decentralized platform must collect participants' offers and make them available to buyers and sellers, and the market algorithm must communicate clearing prices and buyer-seller matchings. These messages must be reliably delivered under strict timing constraints, derived from the deadline by which a trade must clear.
    \item The trading activity shall not compromise the stability of the physical system operation. For example, capacity constraints along any feeder should be respected. Specifically, each feeder is rated for a maximal power capacity. For example, if the feeder capacity is only $10$ kW, then $C_1$ should not consume $30$ kW.
    \item There are a number of parameters of the system that should be made configurable. For example, the number of intervals to look ahead while solving the matching problem is one such parameter. Another parameter is the prediction window for each prosumer. Note that Example~\ref{example:1} required that the prosumers make their offers for future intervals available.
    \item Information such as the amount of energy produced, consumed, bought, or sold by any prosumer should be available only to the Distribution System Operator. All bids and asks as well as the matching thereof should remain anonymous to the other participants. 
    \item The failure of a prosumer or market agent, including any solvers that are required to search for a matching solution, must not compromise the system. Further, there should be mechanisms to ensure that everyone agrees to and conforms to the decisions made by the market mechanism.
\end{itemize}

\section{Related Work}
\label{sec:related}
Implementing a Transaction Management Platforms (TMP) requires a communication architecture, as well as trading mechanisms that provide the capability to match the bids and asks.  Blockchain-based solutions have the potential to enable large-scale energy trading based on distributed consensus systems. However, popular blockchain solutions, such as Bitcoin \cite{Satoshi} and Ethereum \cite{buterin2013ethereum}, suffer from design limitations that prevent their direct application to validating energy trades. %This is primarily due to the lack of additional constraints and checks required, beyond just the transactional integrity check provided by proof-of-work algorithms.
 
For example, Aitzhan and Svetinovic implemented a proof-of-concept platform for decentralized smart-grid energy trading using blockchains, but their system is based on proof-of-work consensus, and they do not consider grid control and stability, or scalability~\cite{aitzhan2016security}. Additionally, there is still the problem of privacy---all transactions in these systems are  public \cite{kosba2016hawk}. 

Most works discussing privacy look at it from the context of smart meters. McDaniel and McLaughlin discuss 
privacy concerns due to energy-usage profiling, which smart grids could
potentially enable~\cite{mcdaniel2009security}. Efthymiou and Kalogridis describe a method for securely anonymizing frequent electrical metering data sent by a smart
meter by using a third party escrow mechanism~\cite{efthymiou2010smart}. Tan et
al.\ study privacy in a smart metering system from an information
theoretic perspective in the presence of energy harvesting and storage
units~\cite{tan2013increasing}. They show that energy harvesting
provides increased privacy by diversifying the energy source, while a
storage device can be used to increase both energy efficiency and
privacy. However, transaction data from energy trading may provide more fine-grained information than smart meter based usage patterns~\cite{Privacy2017}. 

%Majumder et al.\ present an iterative double auction trading mechanism that preserves the participants' privacy~\cite{majumder_efficient_2014}. However, the privacy property pertains to the participants' utility function models, not their identities. 

Existing energy trading markets, such as the European Energy Exchange \cite{EPEX_SPOT_operation} and project NOBEL in Spain, employ the double-auction market mechanism \cite{Ilic12}, which can be implemented to preserve participant privacy. However, typical exchange implementations involve centralized database architectures which constitute single points of failure.
 
Majumder et al.\ present an iterative double auction trading mechanism that preserves the participants' privacy, in particular, it keeps their utility functions private~\cite{majumder_efficient_2014}.
Similarly, Faqiry and Das present an auction mechanism for maximizing social welfare of buyers and sellers (if the supply is small)~\cite{faqiry_transactive_2016}. Their approach also provides some privacy meaning that participants do not reveal their utility functions. By constricting the buyers' utility functions to be convex, the social welfare objective function is maximized when the micro-grid controller objective function, whose goal is to maximize the power sold, is maximized. In the later part of the paper, they consider an approach that discards the privacy maintained during the first phase in order to make  trading fair. In their work, there is no mechanism to check whether the buyer can produce the power they claim they can supply, which could result in instability. The authors also mention in passing that their approach can be implemented as a distributed algorithm, but this was not carried out. 

In contrast, the work presented in this paper is a distributed systems mechanism that considers the problem of a broader definition of privacy, safety, and protection from malicious actors as a combined problem.

\section{\platform Platform}
\label{sec:solution}
\begin{figure}[t]
    \centering
    \includegraphics[width=\columnwidth]{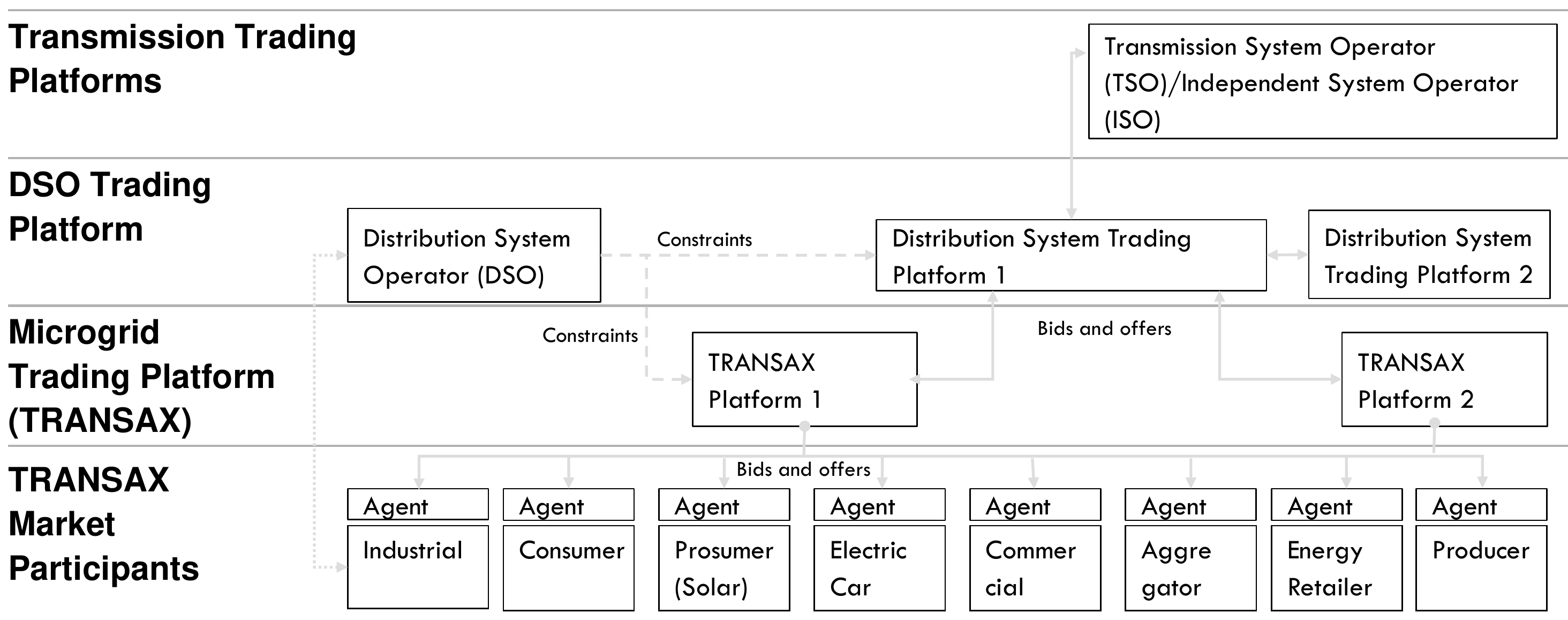}
    \caption{Landscape of a Transactive Energy Systems with TRANSAX.  TSO/ISO are responsible for bulk trading. Multiple instances of smaller scale trading platforms can co-exist at distribution and microgrid level.  }
    \label{fig:architecture}
\end{figure}

\Aron{This section is very long, we might want to split it.}
Next, we describe the solution implemented by the \platform platform. Figure \ref{fig:architecture}
describes the components of this decentralized and distributed platform. An agent runs on a computing node within the premises of each home. In the remainder of this paper, we refer to these agents as ``prosumers'' or ``consumers'' depending upon the context, but they are implemented as one type of entity that can both buy and sell energy. The solvers are nodes responsible for identifying the feasible and optimal trades. The miners  are responsible for reaching consensus on the market solutions using a distributed-ledger based ``smart contract.'' Note that in this architecture, solvers and miners are not centralized. All agents communicate with each other providing offers and accepting the agreed upon market energy transfer schedule. The anonymization mixer implements privacy-preserving mechanisms ensuring that prosumers can remain anonymous to each other. The anonymization module has been explained in a previous paper and is not discussed further here \cite{Laszka17}.

First, we  describe the market problem and then   describe a smart contract solution and a protocol to setup the distributed system. Finally, we describe the distributed architecture of the implementation of the system.  

\subsection{Market Problem}

\begin{table}
\caption{List of Symbols}
\centering
\label{tab:symbols}
\renewcommand*{\arraystretch}{1.2}
\begin{tabular}{|l|p{5.8cm}|}
\hline
Symbol & Description \\
\hline
\multicolumn{2}{|c|}{Microgrid} \\
\hline
$\calF$ & set of feeders \\
\rowcolor{TableRowGray} $C_f^{ext}$ & maximum net power consumption or net power production in feeder $f \in \calF$ \\
$C_f^{int}$ & maximum total power consumption or total power production in feeder $f \in \calF$ \\
\rowcolor{TableRowGray} $\Delta$ & length of time intervals \\
$T_{clear}$ & minimum number of time intervals between the finalization and delivery of a trade \\
\hline
\multicolumn{2}{|c|}{Offers} \\
\hline
\rowcolor{TableRowGray} $\calS_f$ & set of selling offers from feeder $f \in \calF$ \\
$\calB_f$ & set of buying offers from feeder $f \in \calF$ \\
\rowcolor{TableRowGray} $\calS$, $\calB$ & set of all selling and buying offers, resp. \\
$\calS^{(t)}$, $\calB^{(t)}$ & set of selling and buying offers submitted by the end of time interval $t$, resp. \\
\rowcolor{TableRowGray} $E_s$, $E_b$ & amount of energy to be sold or bought by offers $s \in \calS$ and $b \in \calB$, resp. \\
$I_s$, $I_b$ & time intervals in which energy could be provided or consumed by offers $s \in \calS$ and $b \in \calB$, resp. \\
\rowcolor{TableRowGray} $R_s$, $R_b$ & reservation prices of offers $s \in \calS$ and $b \in \calB$, resp. \\
$\calM(s)$, $\calM(b)$ & set of offers that are matchable with offers $s$ and~$b$, resp. \\
\rowcolor{TableRowGray} $I(s, b)$ & $I_s \cap I_b$ \\
\hline
\multicolumn{2}{|c|}{Solution} \\
\hline
$p_{s,b,t}$ & amount of energy that should be provided by $s$ to $b$ in interval $t$ \\
\rowcolor{TableRowGray} $\pi_{s,b,t}$ & unit price for the energy provided by $s$ to $b$ in interval~$t$ \\
$\textit{Feasible}(\calS, \calB)$ & set of feasible solutions given sets of selling and buying offers $\calS$ and $\calB$ \\
\rowcolor{TableRowGray} $\hat{p}_{s,b,t}$, $\hat{\pi}_{s,b,t}$ & finalized trade values \\
\hline
\multicolumn{2}{|c|}{Implementation Parameters} \\
\hline
$T_{predict}$ & prediction window used by prosumers when posting selling and buying offers \\
%$\Delta_s$ & periodicity of solver that matches offers \\
\rowcolor{TableRowGray}  $T_{lookahead}$ &  number of time intervals considered in the future by the solver  \\
%$start$ & simulation start interval \\
%\rowcolor{TableRowGray} $end$ & simulation end interval \\
 $\hat{\Delta}$ &  length of the time step used for simulating the real-interval of length $\Delta$ \\
\hline
\end{tabular}
\end{table}

%In this section, we formulate the problem of matching energy future bids with asks (i.e., offers to buy energy to be delivered in the future with offers to sell energy) in the local energy trading market. Our formulation aims to promote market efficiency by maximizing the amount of energy futures traded within the microgrid, while satisfying microgrid safety requirements. 

Let $\calF$ denote the set of feeders.
For a feeder $f \in \calF$, we let $C_f^{ext}$ denote the maximum amount of power that is allowed to flow into or out of the feeder at any point in time. 
Similarly, we let $C_f^{int}$ denote the maximum amount of power that is allowed to be consumed or produced within the feeder at any point in time.\footnote{In other words, limit $C_f^{ext}$ is imposed on the net production and net consumption of all prosumers in feeder $f$, while limit $C_f^{int}$ is imposed on the total production and total consumption.} 
We assume that time is divided into intervals of fixed length~$\Delta$, and we refer to the $t$-th interval simply as time interval $t$.
For a list of symbols used in the paper, see Table~\ref{tab:symbols}.

The input of the energy trading problem is the set of buying and selling offers posted by the participants.\footnote{Participants may include both prosumers and the DSO. The DSO can shape load and trade energy futures by participating in the energy market the same way as prosumers do.}
For feeder $f \in \calF$, we let $\calS_f$ and $\calB_f$ denote the set of selling and buying offers posted by participants located in that feeder, respectively.\footnote{If the DSO wants to participate in this energy trading market, it can be assigned to a ``dummy'' feeder in the problem formulation.} 
A selling offer $s \in \calS_f$ is a tuple $\left(E_s, I_s, R_s\right)$, where
\begin{itemize}
\item $E_s$ is the amount of energy to be sold,
\item $I_s$ is the set of time intervals in which the energy could be provided,

\item $R_s$ is the reservation price, i.e., lowest unit price for which the participant is willing to sell energy. 
\end{itemize}
Similarly, a buying offer $b \in \calB_f$ is a tuple $\left(E_b, I_b, R_b\right)$, where the values pertain to consuming/buying energy instead of producing/selling, and $R_b$ is the highest price that the participant is willing to pay.
For convenience, we also let $\calS$ and $\calB$ denote the set of all buying and selling offers (i.e., we let $\calS = \cup_{f \in \calF} \calS_f$ and $\calB = \cup_{f \in \calF} \calB_f$).

We say that a pair of selling and buying offers $s \in \calS$ and $b \in \calB$ is \emph{matchable} if
\begin{align}
R_s \leq R_b ~~~~ \textnormal{ and }
~~~~ I_s \cap I_b \neq \emptyset .
\end{align}
In other words, a pair of offers is matchable if there exists a price that both participants would accept and a time interval in which the seller and buyer could provide and consume energy.
For a given selling offer $s \in \calS$,
we let the set of buying offers that are matchable with $s$ be denoted by $\calM(s)$.
Similarly, we let the set of selling offers that are matchable with a buying offer $b$ be denoted by $\calM(b)$.
Further, we let $I(s, b) = I_s \cap I_b$.

A solution to the energy trading problem is a pair of vectors $(\vp, \vpi)$, where
\begin{itemize}
\item $p_{s,b,t}$ is a non-negative amount of power that should be provided by the seller $s \in \calS$ and consumed by the buyer $b \in \calM(s)$ in time interval $t \in I(s, b)$.\footnote{We require the both the seller and buyer to produce a constant level of power during the time interval.}

\item $\pi_{s,b,t}$ is the unit price for the energy provided by seller $s \in \calS$ to buyer $b \in \calM(s)$ in time interval $t \in I(s, b)$.
\end{itemize}

A pair of vectors $(\vp, \vpi)$ is a feasible solution to the energy trading problem if it satisfies the following constraints:
\begin{itemize}
\item The amount of energy sold or bought from each offer is at most the amount of energy offered:
\begin{align}
\forall s \in \calS: ~ \sum_{b \in \calM(s)} \sum_{t \in I(s,b)} p_{s,b,t} \cdot \Delta \leq E_s \label{eq:constrEnergyProd} \\
\forall b \in \calB: ~ \sum_{s \in \calM(b)} \sum_{t \in I(s,b)} p_{s,b,t} \cdot \Delta \leq E_b \label{eq:constrEnergyCons} 
\end{align}
\item The amount of power flowing into or out of each feeder is below the safety limit in all time intervals:

\begin{align}
\forall f \in & \, \calF, t: \nonumber \\ 
& \left( \sum_{s \in \calS_f} \sum_{b \in \calB} p_{s,b,t} \right) - \left( \sum_{b \in \calB_f} \sum_{s \in \calS} p_{s,b,t} \right) \leq C_f^{ext} \label{eq:constrExtProd} \\
\forall f \in & \, \calF, t: \nonumber \\
& \left( \sum_{s \in \calS_f} \sum_{b \in \calB} p_{s,b,t} \right) - \left( \sum_{b \in \calB_f} \sum_{s \in \calS} p_{s,b,t} \right) \geq -C_f^{ext} \label{eq:constrExtCons} 
\end{align}
\item The amount of energy consumed and produced within each feeder is below the safety limit in all time intervals:
\begin{align}
\forall f \in \calF, t: ~ \sum_{b \in \calB_f} \sum_{s \in \calS} p_{s,b,t} \leq C_f^{int} \label{eq:constrIntCons} \\
\forall f \in \calF, t: ~ \sum_{s \in \calS_f} \sum_{b \in \calB} p_{s,b,t} \leq C_f^{int} \label{eq:constrIntProd} 
\end{align}
\item The unit prices are between the reservation prices of the seller and buyer:
\begin{align}
\forall s \in \calS, b \in \calM(s), t \in I(s,b): ~ R_s \leq \pi_{s,b,t} \leq R_b \label{eq:constrPrice} 
\end{align}
\end{itemize}

The objective of the energy trading problem is to maximize the amount of energy traded.
Formally, an optimal solution to the energy trading problem is
\begin{align}
\label{optimalproblem}
\max_{(\vp,\vpi) \, \in \textit{ Feasible}(\calS, \calB)} ~
\sum_{s \in \calS} \sum_{b \in \calM(s)} \sum_{t \in I(s,b)} p_{s,b,t} \, , 
\end{align}
where $\textit{Feasible}(\calS, \calB)$ is the set of feasible solutions given selling and buying offers $\calS$ and $\calB$ (i.e., set of solutions satisfying Equations~\eqref{eq:constrEnergyProd} to \eqref{eq:constrPrice} with $\calS$ and $\calB$).

\subsubsection{Linear-Programming Solution:} We can solve the basic energy trading problem efficiently by formulating it as a linear program.
First, create real-valued variables $p_{s,b,t}$ and $\pi_{s,b,t}$ for each $s \in S, b \in \calM(s), t \in I(s, b)$.
Then, the following reformulation of the matching problem is a linear program:
\begin{equation}
\max_{\vp,\vpi}
\sum_{s \in \calS} \sum_{b \in \calM(s)} \sum_{t \in I(s,b)} p_{s,b,t} \label{eq:linProgObj}
\end{equation}
subject to Equations~\eqref{eq:constrEnergyProd} to \eqref{eq:constrPrice} and
\begin{equation}
\vp \geq \vect{0} \text{ and } \vpi \geq \vect{0} .
\end{equation} 

\subsubsection{Trade Finalization} Equation \eqref{optimalproblem} formulates the problem considering a single ``snapshot'' of all offers across all time intervals. However, in practice, prosumers may submit new offers at any time, resulting in continuously evolving sets of offers.
Consequently, optimal solutions to Equation \eqref{optimalproblem} may have to be found repeatedly as new offers are submitted, resulting in a series of evolving solutions.
This presents a problem since prosumers need to know in advance what the ``final'' solution for a certain time interval is in order to be able to actually schedule energy production or consumption for that interval. 
Further, preventing ``last-minute'' changes can be crucial for safety and stability since it allows the DSO to prepare for satisfying energy demand that cannot be met locally.

As the set of submitted offers grows, the optimal solution to the energy trading problem may change, and the optimal value of each $p_{s,b,t}$ may vary.
While each change can increase the amount of energy traded, the \emph{trade values} $p_{s,b,t}$ and $\pi_{s,b,t}$ need to be \emph{finalized} at some point in time.
At the very latest, values for interval $t$ need to be finalized by the end of interval $t - 1$; otherwise, participants would have no chance of actually delivering the trade.
We now extend the energy trading problem to consider a growing set of offers and a time constraint for finalizing trades.
Our approach finalizes a minimum set of trades in each interval, which maximizes efficiency while providing safety.

We assume that all trades for time interval $t$ (i.e., all values~$p_{s,b,t}$ and $\pi_{s,b,t}$) must be finalized by the end of time interval $t - T_{clear}$, where $T_{clear}$ is a positive integer constant, which is set by the operator. In practice, the value of $T_{clear}$ must be chosen taking into account both physical constraints (e.g., how long it takes to turn on a generator) and communication delay (e.g., some participants might learn of a trade with delay due to network disruptions).

We let $\hat{p}_{s,b,t}$ and $\hat{\pi}_{s,b,t}$ denote the finalized trade values, and we let $\calB^{(t)}$ and $\calS^{(t)}$ denote the set of buying and selling offers that participants have submitted by the end of time interval $t$.
Then, the system takes the following steps at the end of each time interval $t$:
\begin{itemize}
\item Find an optimal solution $(\vp^*, \vpi^*)$ to the extended energy trading problem:
\begin{equation}
\label{eq:ext_prob}
\max_{(\vp, \vpi) \, \in \textit{ Feasible}(\calS^{(t)}\!,\, \calB^{(t)})} \sum_{s \in \calS^{(t)}} \sum_{b \in \calM(s)} \sum_{\tau \in I(s,b)} p_{s,b,\tau}
%\max_{\substack{(\vp, \vpi) \, \in \textit{ Feasible}(\calS^{(t)}\!,\, \calB^{(t)}) \\ \text{subject to} \\ \forall \tau \leq t - T_{clear}: ~ \hat{p}_{s,b,t} = p_{s,b,t} \wedge \hat{\pi}_{s,b,t} = \pi_{s,b,t}}}
\end{equation}
subject to
\begin{align}
\forall \, \tau < t + T_{clear}\!: ~ & p_{s,b,\tau} = \hat{p}_{s,b,\tau} \\ %~\wedge~ \pi_{s,b,t} = \hat{\pi}_{s,b,t} .
& \pi_{s,b,\tau} = \hat{\pi}_{s,b,\tau} 
\end{align}
\item Finalize trade values for time interval $t + T_{clear}$ based on the optimal solution $(\vp^*, \vpi^*)$:
\begin{align}
& \hat{p}_{s,b,t + T_{clear}} := p^*_{s,b,t + T_{clear}} \\ 
& \hat{\pi}_{s,b,t + T_{clear}} := \pi^*_{s,b,t + T_{clear}} 
\end{align}
\end{itemize}

The problem in Equation~\eqref{eq:ext_prob} can also be reformulated as a linear program similarly, by considering $\calS^{(t)}$, $\calB^{(t)}$, $\hat{\vp}$, $\hat{\vpi}$, and the additional constraints.

\subsection{Market Solver}
The role of the market solver is to periodically solve the linear program mentioned above as the offers stream in. To address the trustworthiness challenge, we implement a blockchain based solution as discussed previously. However, since computation is relatively expensive on blockchain-based distributed platforms, solving the energy trading problem using a block\-chain-based smart contract would not be scalable in practice.
In light of this, we adopt a \emph{hybrid implementation approach}, which we introduced in earlier~\cite{eisele2018solidworx}, to transactive energy systems.
The hybrid approach combines the trustworthiness of blockchain-based smart contracts with the efficiency of more traditional computational platforms. The key idea of our hybrid approach is to (1) use a high-performance computer to solve the computationally expensive linear program \emph{off-blockchain}\footnote{We use CPLEX \cite{cplex2009v12} as the MILP solver engine in \platform.} and then (2) use a smart contract to record the solution \emph{on the blockchain}.

\subsubsection{Blockchain-based Smart Contract}
We implemented a smart contract\footnote{Source code is available upon request.} that   verifies the feasibility of each  solution $(\vp, \vpi)$ submitted by an off-blockchain solver.  If the solution is feasible, then the contract computes the value of the objective function and compares it to the objective value of previously submitted solutions.
The contract always keeps track of the best feasible solution submitted so far, which we call the \emph{candidate solution}. At the end of each time interval $t$, the contract finalizes the trade values for interval $t + T_{clear}$ based on the candidate solution.\footnote{If no solution has been submitted to the contract so far, which might be the case right after the trading system has been launched, $\vp = \vect{0}$ may be used as a candidate solution.}

This simple functionality achieves a high level of security and reliability.
Firstly, it is clear that an adversary cannot force the contract to finalize trades based on an unsafe (i.e., infeasible) solution since such a solution would be rejected.
Similarly, an adversary cannot force the contract to choose an inferior solution instead of a superior one.
In sum, the only action available to the adversary is proposing a superior feasible solution, which would actually improve energy trading in the microgrid.

The contract is also reliable and can tolerate temporary disruptions in the solver or the communication network.
Notice that any solution $(\vp, \vpi)$ that is feasible for sets $\calS$ and $\calB$ is also feasible for supersets $\calS' \supseteq \calS$ and $\calB' \supseteq \calB$.
As the sets of offers can only grow over time, the contract can use a candidate solution submitted during time interval $t$ to finalize trades in any subsequent time interval $\tau > t$.
In fact, without receiving new solutions, the difference between the amount of finalized trades and the optimum will increase only gradually:
since the earlier candidate solution can specify trades for any future time interval,
the difference is only due to the offers that have been posted since the solution was found and submitted.

%In our hybrid approach, we implement a relatively simple algorithm in our smart contract, which can (1) check if a solution $(\vp, \vpi)$ is feasible and (2) evaluate the objective function for a given solution (see Equation~\eqref{eq:linProgObj}).

\subsubsection{Solver}
We complement the smart contract with an efficient linear programming solver, which can be run off-blockchain, on any capable computer (or multiple computers for reliability).
The solver is run periodically to find a solution to the energy trading problem based on the latest set of offers posted. 
Once a solution is found by the solver, it is submitted to the smart contract in a blockchain transaction.
Note that if new offers have been posted since the solver started working on the solution, the contract will still consider the solution to be feasible. 
This is again due to any feasible solution for sets $\calS$ and $\calB$ also being feasible for supersets $\calS' \supseteq \calS$ and $\calB' \supseteq \calB$.

From the perspective of the solver, being able to submit multiple solutions to the contract for the same problem has many advantages.
For example, it allows the linear programming solver to be run as an anytime algorithm.
Further, we can allow multiple---potentially untrusted---entities to try to solve the problem and submit solutions, since the smart contract will always choose the best feasible one.
This is especially important in microgrids where a trusted third party is not guaranteed to always be present.
In such settings, prosumers can be allowed to volunteer and provide solutions to the energy trading problem.%
\footnote{Of course, each prosumer will try to submit a solution that favors the prosumer. However, the submitted solution still needs to be superior with respect to the optimization objective, which roughly corresponds to social utility. Hence, each prosumer is incentivized to improve social utility by submitting a superior solutions that favors the prosumer. We leave the analysis of these incentives for future work.}
Thereby, we enable finding solutions in an efficient and very flexible manner, while reaping the benefits of smart contracts, such as auditability and trustworthiness.

% \begin{figure}[t]
%     \centering
%     \includegraphics[width=\columnwidth]{figs/memory.png}
%     \caption{Memory consumption when solving a single interval of the simulation (interval 40) for various lengths of the lookahead window.}
%     \label{fig:memory}
% \end{figure}

\subsubsection{Solver Lookahead Window}
\label{sec:lookahead}
%
%Suppose that the offers made by prosumers $P_1, P_2$ and consumer $C_1$ from Example \ref{example:1} are $P_1$: \{OfferID:1, ProsumerID:1, Start:48, End:48, Quantity:10\}, 
% $P_2$ : \{OfferID:2, ProsumerID:2, Start:48, End:49, Quantity:30\}, and 
%$C_1$ with \{OfferID:3, ProsumerID:3, Start:48, End:48, Quantity:-30\}, and 
%\{OfferID:4, ProsumerID:3, Start:49, End:49, Quantity:-10\}. Then, if the lookahead window  of the solver is 0, it will identify a solution with the extremes being (1) \{b: 3, s: 2, t: 48, p: 30\} matching offer 3 and 2 for 30 kW, and (2) \{b: 3, s: 2, t: 48, p: 20\} matching offer 3 and 2 for 20 kW and \{b: 3, s: 1, t: 48, p: 10\} matching offer 3 and 1 for 10 kW. Note that the second solution leaves the offers:  \{2, 2, 48, 49, 10\} (the residual left for prosumer $P_2$), and \{4, 3, 49, 49, -10\}, the offer from consumer $C_1$. The second extreme allows overall 40 kW to be transferred compared to 30 kW total in the first. However, picking the second alternative requires the solver to consider both offers from interval 48 and 49 while computing solutions, effective requiring a lookahead window of 1. 
%

Since the energy trading problem (i.e., Equation~\eqref{eq:ext_prob}) can be formulated as a linear program, we can solve it efficiently, that is, in polynomial time.
However, as the number of offers and the time intervals that they span increases, the number of variables $\{p_{s,b,t}\}$ may grow prohibitively high, which makes solving the trading problem very challenging in practice.
A key observation that helps us tackle this challenge is that even though consumers and prosumers may post offers whose latest intervals are in the far future (i.e., for an offer $s$, the latest interval may be $\max I_s \gg t$, where $t$ is the current interval), a solver only needs to consider a few intervals ahead of the finalization deadline.
Indeed, we have observed that considering intervals in the far future has little effect on the optimal solution for the interval that is to be finalized next.

Consequently, for practical solvers, we introduce a lookahead window $T_{lookahead} \geq T_{clear}$ that limits the intervals that need to be considered effectively: for any $\hat{t} > t + T_{lookahead}$, we set $p_{s,b,\hat{t}} = 0$, where $t$ is the current interval.
By ``pruning'' the set of fee variables, we can significantly improve the performance of the solver with negligible effect on solution quality.
Figure \ref{fig:loahNrgMEM} shows the memory usage of the solver (in time interval $t = 80$) and the energy traded, while varying the lookahead-window length $T_{lookahead}$. 

Similar to memory, the lookahead parameter also impacts the CPU utilization of the solver. Thus, as a practical matter, we implemented a hierarchical controller to automatically adjust the lookahead window in \platform solvers using resource limit callbacks, which we will describe in Section~\ref{sec:implementation}. The top-level controller sets the maximum lookahead value based on the available memory. The low-level controller sets the lookahead to a value between $T_{clear}$ and the upper bound. The asynchronous architecture of \platform enables multiple solvers to operate simultaneously and compete in providing a better matching solution, while obeying the limits imposed by available resources. This ensures that the solvers can be run on edge computing nodes in a community where other applications might also be co-hosted.

%\textcolor{red}{To set the value of $T_{lookahead}$, we implement a proportional controller which sets it to a value between 0 and an upper bound $T_{max}$. $T_{max}$ can be adjusted by the memory resource limit violation handlers in Table \ref{tab:events}. The set point of the controller is a solve time of 0.5 seconds, and the error is measured each time the solver finds a new solution to post. The controller could alternatively control $T_{lookahead}$ based on other resources, RAM, or CPU utilization.}

\subsubsection{Other Parameters}
In addition to the lookahead window $T_{lookahead}$, our implementation can also be configured with parameters that control the prosumers and the speed of the simulation. 
%The simulation parameters can be found in Table \ref{tab:symbols} under \textit{Implementation Parameters}.
The solver  operates as a periodic process, % (with a period $\Delta_s$), 
waiting on information from the smart contract about all the offers that have been posted. In our implementation, the prosumers also operate periodically, submitting their offers and bids to the smart contract in every interval. In a given interval, our prosumer implementation provides offers for up to $T_{predict}$ intervals in the future (including the current interval), where $T_{predict}$ is a parameter of the prosumer. We require that $T_{predict} > 1$ because we need at least one interval prediction for trading energy futures. %$T_{clear}$ as the readers may recall is the parameter of the system forcing a finalization deadline. Thus in interval $t$, we must have finalized up to intervals $t+T_{clear}$. The solver has a parameter $T_{lookahead}$, which decides how many intervals from the future the solver will use to solve the matching problem. 
Finally, during our experiments, we may speed up the simulation by letting the real-time length of the time interval be $\hat{\Delta} < \Delta$, but keeping the theoretical length of the interval at~$\Delta$. Note that $\hat{\Delta}$ is the amount of real time passed in the simulation before proceeding to the next interval. This allows us to speed  up the experiments without compromising our results since running the system slower would be easier.

\begin{figure} [t]
    \centering
    \includegraphics[trim=.9cm 7cm 5.9cm .9cm, width=\columnwidth]{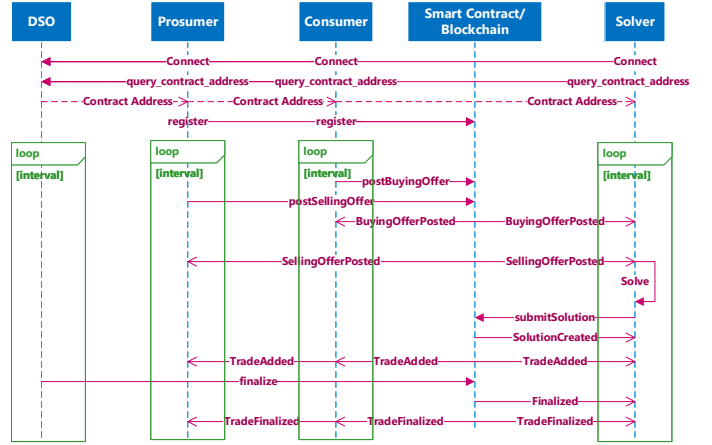}
    \caption{Interaction diagram of \platform components.}
    \label{fig:wf}
\end{figure}

\subsection{\platform Protocol} 

\begin{figure} [t]
%source: https://docs.google.com/spreadsheets/d/1EoQKCElVXEhQFYyZa_DSOTMs6ox_NBGuxqHz0NxtSew/edit?usp=sharing
    \centering
% \vspace{-0.1in}
%    \makebox[\linewidth]{
    % \includegraphics[width=1.0\linewidth]{figs/loahNrgMEM.png}
    \includegraphics[width=1.0\linewidth]{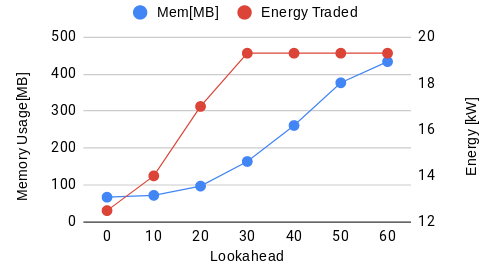}
%    }
    %\vspace{-0.2em}
    \caption{Memory consumption and Energy traded during a single interval of the simulation (interval 80) for various values of $T_{lookahead}$.}
    \label{fig:loahNrgMEM}
    \vspace{-0.12in}
\end{figure}
 
As illustrated in Figure \ref{fig:wf}, when a participant (i.e., prosumer, consumer, or solver) receives the address of the smart contract, they submit a transaction to register themselves in the blockchain. After some time, the transaction is mined and triggers an event (e.g., \texttt{ProsumerRegistered}) notifying the participant that it can begin posting offers. When the participants contact the DSO, the response contains time information allowing the participants to determine the earliest interval for which the blockchain is accepting  bids and solutions. Any bid or solutions that contains an end time after that interval is ignored. The interval returned by the DSO is some number of intervals ahead of the current interval of the microgrid, since the power schedule must be determined before the time of actuation. The length of an interval in the case studies described here is 1 minute. For live deployments, the value can be configured by the system integrator. At the start of each interval, prosumers submit relevant bids to the blockchain. After trades have been added to the blockchain, the solver receives the \texttt{OfferPosted} event and will attempt to find a valid matching between bids and requests. The solver has a solving interval and attempts to find a better solution during each one. This continues until the DSO submits a \texttt{finalize} transaction to the blockchain, which triggers a \texttt{TradeFinalized} event. This event causes the solver  to update its interval and begin working to find a solution for the next interval. The prosumers also receive this event informing them of the power they are expected to produce/consume during that interval when it arrives.  Two concepts explained below are critical to this protocol

%\Aron{This paragraph has been commented out, but I think that some parts of it could be added back before the preceding paragraph.}
%Before discussing the distributed system architecture, constraints, and setup, we provide a quick overview of how the prosumers, consumers, and market solvers interact (Figure \ref{fig:wf}). Every agent receive a set of anonymous identifiers and the address of a miner (any miner running the smart contract) from the DSO.  Thereafter, every agent registers with the smart contract. The registration information requires each agent to specify the feeder to which they belong, which is certified by a signature from the DSO. This is required to ensure that feeder-level safety constraints can be correctly applied. These registrations can happen asynchronously, allowing a prosumer to  join at any time, even long after trading has commenced.

\subsubsection{Offers} The middleware platform on which \platform is build (described in next section) solves the problem of time synchronization, enabling all agents to correctly know the current interval and current time. Thus, the solvers and prosumers do not need to synchronize independently and can keep track of offers and intervals. The distributed ledger acts as a shared database and provides a notification log of events (e.g., submission of a new offer). In \platform, these offers are sent as a structure of following form \texttt{\{PROSUMER ID, START INTERVAL, END INTERVAL, ENERGY QUANTITY\}}.
Agents and solvers listen for these events and perform actions based on them.

%Solvers keep an internal clock and based on the parameters of (a) how many future intervals to consider, and (b) current interval computes solutions, which are then evaluated by the smart contract. The miners work to obtain consensus and register the decision of the smart contract (identifying the best match for an interval) into the distributed ledger. The smart contracts finalize the solution based on a signal from DSO (generated based on the setting $T_{clear}$  -- see table \ref{tab:symbols}). 

\subsubsection{Trade Finalization}
Finalization of an interval means that the smart contract will not accept any more changes to the solution for that interval. 
Prosumers are notified of finalized trades using \texttt{TradeFinalized} events, which communicate matches as structures of the following form \texttt{\{BUY OFFER ID, SELL OFFER ID, INTERVAL, POWER\}}. This  enables prosumers to act according to the solution of the energy trading problem since they know the identifiers of their offers and can filter on finalized trades. Note that even though we do not discuss penalizing prosumers who do not conform to the solution, it is straightforward to do this since the DSO can associate prosumers with their anonymous identifiers, and all offers and trades are permanently recorded on the ledger. 
Finally, the DSO can combine the recorded trades with actual power consumption and production values measured by electricity meters in order to bill prosumers (e.g., every month).

\subsection{\platform Implementation and Resilience}
\label{sec:implementation}
\platform is implemented as an application in the Resilient Information Architecture Platform for Smart Grid (RIAPS) \cite{Scott2017ICCPS}.  A RIAPS system is a collection of computing nodes, which are connected to power system sensors and actuators over a variety of interfaces, e.g. Modbus/UART, etc.  Each RIAPS computing node executes a collection of platform services which run with the highest privileges on the system. These services are for discovering the other components in the distributed system (\texttt{riaps\_disco}), for remotely deploying application (\texttt{riaps\_deplo}) and for providing the computing nodes with a synchronized time \cite{riapsrsp}, which is critical for the correct operation of \platform. Control nodes are responsible for installing and removing distributed applications on these nodes. Each RIAPS application is a collection of \textit{actors}, which are assembled from reusable \textit{components}. 

\subsubsection{Fault Model} 
We extended RIAPS with additional capabilities to monitor and mitigate failures across three layers: physical and device level, platform services level, and application level. Typically, physical electrical-system architectures are designed with $N-1$ criterion, i.e., they have redundancy to tolerate the failure of any single physical device. Fault management at services level is implemented by the combination of a distributed hash table, which maintains information about all actors, and Zero MQ Zyre \cite{hintjens2013zeromq}, an open-source framework for proximity-based peer-to-peer applications.  Further, we extended RIAPS to provide resource management.  These features are important to ensure that \platform actors can work on remote nodes within the limits of available resources. These limits are enforced via the use of the cgroups interface, watchdogs, and custom zeroMQ pair connections in RIAPS. To support application-specific failure mitigation, RIAPS provides callbacks (see Table \ref{tab:events}), which can be implemented by the component developers, in this case the \platform components. For example, we used the $handleCPULimit$ to implement the controller for the lookahead window described in Section \ref{sec:lookahead}.

\begin{table}
\caption{Callback handlers implemented in RIAPS for providing application specific mitigation actions. These handlers respond to failures at three levels: application code, RIAPS services and Physical devices.}
\centering
\label{tab:events}
\renewcommand*{\arraystretch}{1.2}
\resizebox{\columnwidth}{!}{%
\begin{tabular}{|l|l|l|}
\hline
Source & Handler & Event \\
\hline
% \multicolumn{2}{|c|}{Microgrid} \\
% \hline
App & $handleCPULimit$ & App CPU usage exceeds threshold \\
\rowcolor{TableRowGray} App & $handleMemLimit$ & App Memory usage exceeds threshold\\
App & $handleSpcLimit$ & App Disk usage exceeds threshold\\
\rowcolor{TableRowGray}  App & $handleNetLimit$ & App Network usage exceeds threshold \\
Phy & $handleNICStateChange$ & RIAPS node network status has changed \\
\rowcolor{TableRowGray}  App, service, Phy & $handlePeerStateChange$ & Clean ingress/egress to network of peer \\
App, service & $handleDeadline$ & App function duration exceeds threshold \\ 
\hline
\end{tabular}}
\end{table}

\subsubsection{Mining Considerations} In the current implementation of \platform, we use a private Ethereum network as the distributed ledger. To speed up the consensus protocol, we reduce the the difficulty of the cryptographic puzzle solved for proof-of-work consensus. For larger systems, the proof-of-work consensus may be replaced by, e.g., proof-of-stake, for scalability.

\section{Case Study}
\label{sec:casetudy}

 We consider a collection of load traces recorded  from a microgrid in Germany, containing $102$ homes ($5$ producers, $97$ consumers) across $11$ feeders. We show the nominal execution of the system as well as its resilience capabilities by illustrating execution under resource constraints and actor failures.

 %Figure~\ref{fig:feeder} shows the distribution of prosumers across the feeders, and the overcurrent relays.
%
 %Table \ref{tab:prosumers}
 %We use $\Delta =15$ minute intervals, resulting in a total of $96$ intervals across the whole day.   Figure~\ref{fig:profile} shows the total production and consumption across this microgrid. The horizontal axis shows the starting time for each of the $96$ intervals. 
% Since the dataset does not include prices, we assume reservation prices to be uniform in our experiments, and focus on studying the amount of energy traded and the performance of the system.
%
%This simulations for our case study were performed on a virtual machine configured with 8GB of RAM and 2 cores of an i7-6700HQ processor. 

% \begin{figure} [ht]
%     \centering
% % \vspace{-0.1in}
% %    \makebox[\linewidth]{
%     \includegraphics[width=1.0\linewidth]{figs/arch-sim1.png}
% %    }
%     %\vspace{-0.2em}
%     \caption{The architecture of the market-only simulation}
%     \label{fig:arch-sim1}
%     \vspace{-0.12in}
% \end{figure}

\subsubsection{Nominal Evaluation} 
\begin{figure} [t]
    \centering
% \vspace{-0.1in}
%    \makebox[\linewidth]{
    \includegraphics[width=1.0\linewidth]{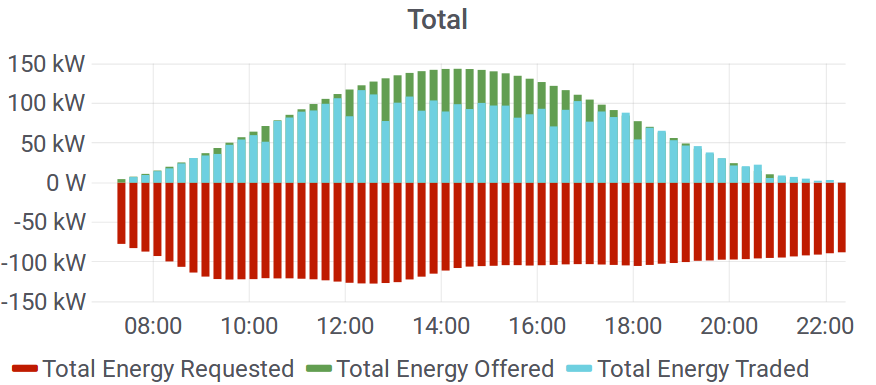}
%    }
    %\vspace{-0.2em}
    \caption{Total energy production capacity (green), and energy demand (red) for each interval, as well as the total energy traded in each interval (blue) while subject to constraints. $C_{ext}=2$MW, $C_{int}=2.5$MW.}
    \label{fig:total}
    \vspace{-0.12in}
\end{figure}

At this scale (102 prosumers), the current implementation was able to match offers during each simulation interval with $T_{lookahead}=5$. The system-wide trading results can be seen in Figure \ref{fig:total}. Each bar is a 15 minute interval. A green bar is the sum of all energy selling offers during that interval. A red bar is the sum of all energy buying offers during that interval. The blue bars are overlayed on the green bars, showing the total energy traded during that interval.  

\begin{figure} [t]
    \centering
% \vspace{-0.1in}
%    \makebox[\linewidth]{
    \includegraphics[width=1.0\linewidth]{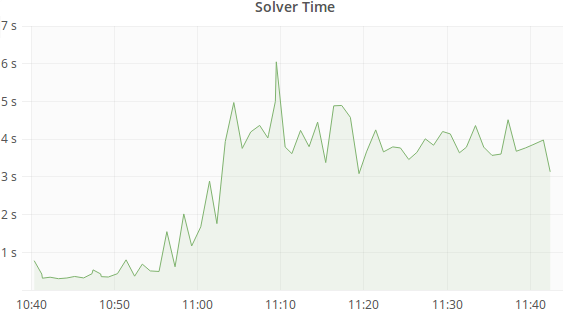}
%    }
    %\vspace{-0.2em}
    \caption{Time taken by the solver to finish each time it runs. The solver runs every 5 seconds. 
    %The times plotted are the actual clock time, we can see that the simulation ran for about an hour. We also note that at 11:00am 20 minutes into the simulation which is interval 48, corresponding to noon in the simulation,  the solve time begins to increase. This is because some of the selling offers are valid for a range of future intervals because of battery storage rather than requiring immediate consumption. 
    }
    \label{fig:solve-time}
    \vspace{-0.12in}
\end{figure}

Early in the simulation, the buying offers exceed the selling offers. Then, as solar generation increases, the selling offers exceed the buying offers. The excess may be stored in batteries for use in future intervals, which increases the complexity of the MILP problem since offers can be matched across multiple intervals. Figure \ref{fig:solve-time} shows evidence of this fact as we see an increase in solver time when selling offers exceed buying offers, around 11:00am. The increasing solver time is the result of increasing problem complexity, which is correlated with the number of variables and constraints in a problem. Some intervals and the corresponding numbers of variables and solve times are shown in Table \ref{tab:solveTimes}. Again, we see that as the selling offers exceed the buying offers, complexity increases, which results in increased solve time. These results provide insight into how the solver scales.

\begin{table}
\caption{Solving times as problem complexity increases}
% Source : https://docs.google.com/spreadsheets/d/1PGu9CTGX5xo3JjRP_bYk3oCDUdW0mfatJMNzXEVZC5Y/edit?usp=sharing
\centering
\label{tab:solveTimes}
\renewcommand*{\arraystretch}{1.2}
\resizebox{\columnwidth}{!}{%
\begin{tabular}{|l|l|l|l|l|}
\hline 
Real Time & Simulation Time & Variables & Constraints & Solving Time \\
\hline
\rowcolor{TableRowGray} 10:52am & 10:00am & 2910 & 750 & 0.891s \\
                        11:00am & 12:00am & 6984 & 762 & 1.612s \\
\rowcolor{TableRowGray} 11:04am & 1:00pm  & 17751 & 778 & 4.35s \\
\hline
\end{tabular}}
\end{table}

The scalability of \platform is limited by the number of transactions that the distributed ledger supports, as well as the complexity of the MILP problem determined by the number of constraints and variables. 
% \textcolor{red}{According to \cite{?} Ethereum currently supports x transactions per second, and they are working to enhance scalability}. \Aron{We might want to omit this number since Ethereum is currently way too slow for us. We can just note that researchers and industry are actively working on significantly improving the transaction performance of Ethereum and other blockchains.}
Additionally, \platform is able to scale by reducing the number offers in a particular interval by suggesting that solvers reduce their lookahead window. 

The number of trades that are made in a day depends on the system parameters. In two experiments, we modulated the power flow constraints $C_{ext}$ and $C_{int}$. The result of this can be seen in Table~\ref{tab:stats}. In both cases, the total buying and selling offers remained constant, only the amount of power that was permitted to flow was changed. Changing the constraints increased the power traded by 1.4MW, thus reducing unused energy by 31\%. In light of this, the efficiency of the platform is primarily dependent on the offers that prosumers make and the system constraints. 

\begin{table}[t]
\caption{Monetary impact of trades subject to constraints $C_{ext}$ and $C_{int}$, given \$0.12/KWh, and total production and demand for energy of $\sim$4.5MW and $\sim$8.3MW, respectively. }
% Source : https://docs.google.com/spreadsheets/d/1PGu9CTGX5xo3JjRP_bYk3oCDUdW0mfatJMNzXEVZC5Y/edit?usp=sharing
\centering
\label{tab:stats}
\renewcommand*{\arraystretch}{1.2}
\resizebox{\columnwidth}{!}{%
\begin{tabular}{|l|l|l|l|}
\hline 
$C_{ext}$, $C_{int}$  & Traded & Unused Energy & Unmet Demand \\
\hline
\rowcolor{TableRowGray} 20kW, 25kW & 2.288MW& 50\% (\$270.73) & 73\% (\$722.90)\\
2MW, 2.5MW & 3.668MW & 19\% (\$105.18) & 56\% (\$557.35) \\
\hline
\end{tabular}}
\end{table}

\subsubsection{Resource Limit Evaluation} 
\begin{figure} [t]
    \centering
% \vspace{-0.1in}
%    \makebox[\linewidth]{
    \includegraphics[width=1.0\linewidth]{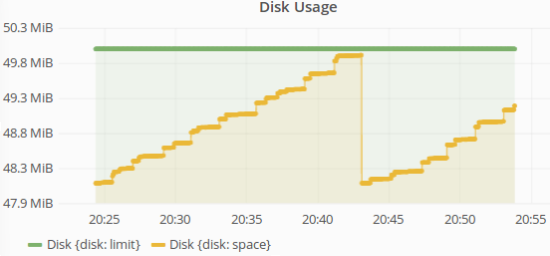}
%    }
    %\vspace{-0.2em}
    \caption{The application disk usage grows from 48MB to the disk limit of 50MB, at which point the handler fires and allows the application to take corrective actions. }
    \label{fig:diskLimit}
    \vspace{-0.12in}
\end{figure}

In this section, we show resource-limit monitoring and mitigation for disk usage and CPU usage. In Figure \ref{fig:diskLimit}, we set the disk storage limit for the prosumer to 50 MB. When the limit is reached,  $handleSpcLimit$ is triggered (see Table \ref{tab:events}), which forces the prosumer to rotate the logs.

% The call back logic implemented in the prosumer actor 

% The disk space limit specifies the maximum space an application is permitted to use. If the limit is reached  $handleSpcLimit$ (see Table \ref{tab:events}) is triggered. In Figure \ref{fig:diskLimit}, the limit was set to 50MB, shown by the green points. The application size is tracked by the yellow points. The application size grows as the solver writes logs which consume additional disk space. When the limit is reached the handler fires and in this case is used to delete the log file.\Aron{Is deleting log files a novel contribution?}

\begin{figure} [t]
    \centering
% \vspace{-0.1in}
%    \makebox[\linewidth]{
    % \includegraphics[width=1.0\linewidth]{figs/cpuLimitHandler.png}
    \includegraphics[width=1.0\linewidth]{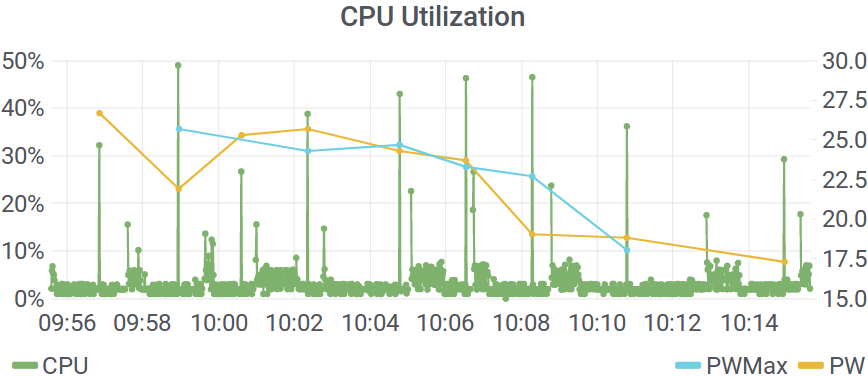}
%    }
    %\vspace{-0.2em}
    \caption{$handleCPULimit$ reduces the maximum lookahead when the CPU limit is crossed.}
    \label{fig:cpuLimit}
    \vspace{-0.12in}
\end{figure}

To show the effect of the CPU resource constraint, we refer back to Section \ref{sec:lookahead}. The actions of the top-level controller can be seen in Figure \ref{fig:cpuLimit} as the yellow dots. When the solver consumes more than 30\% \footnote{These instances can be seen in the \textit{\% CPU Utilization} plot of Figure \ref{fig:cpuLimit}.} of the CPU, the top-level controller reduces the maximum value that the low-level controller may set. We see that over time, the maximum value decreases and the lookahead value (green dots) stays below the maximum. The low-level controller sets the value of the lookahead window, and its influence is shown by the green dots. The low-level controller is implemented as a proportional controller which monitors the solve time and has a solve-time set point of 0.5 seconds. This value was chosen for testing purposes only. The memory controller (not pictured) uses the same high-level control (when the threshold is crossed, it reduces the upper bound) and the same low level control.

Figure \ref{fig:loahNrgMEM} demonstrates how \platform can adapt to variation in the problem complexity. It shows the trade-off between resource consumption (memory) and trading efficiency during interval 80 (8:00pm) as the lookahead window varies. Trading efficiency stops increasing with a lookahead of 30, since interval 50 (12:30pm) is when selling offers become larger than buying offers, and thus the interval in which prediction becomes beneficial. 

\subsubsection{Failure Evaluation}
Since \platform is decentralized, there may be any number of solvers communicating asynchronously with the smart contract, being notified of trades and posting potential solutions. Thus, if any given solver fails, the system will continue unimpeded as long as other solvers are operational. The event $handlePeerStateChange$ has been implemented in the platform, and it is triggered when a node in the network fails or if it is disconnected for any reason. All the peers in the network may receive this message and take actions as appropriate. For example, in the transactive energy case, if a node goes down, it may be unable to provide the energy it was supposed to, and so its trades should be removed. During the testing of the fault-tolerance features, node failure was detected on average in $0.14$ seconds. The peers (other actors in the \platform application) were notified that the node has recovered $1.88$ seconds after the failure, and the node was able to fully reactivate in $6.52$ seconds on average. 

%after the initial failure after 1.88 seconds. 

% KILL PROCESS: 1532148394.986511
% Live: 1532148401.5066464

% WARNING:04:46:35,126:[14799]:Recorder:peer 1668CBFD8AA4172AAB68A2A32C6A6E94 is off at: 1532148395.1262708
% WARNING:04:46:36,863:[14799]:Recorder:peer 1668CBFD8AA4172AAB68A2A32C6A6E94 is on at: 1532148396.863491

\section{Conclusions and Discussions}
\label{sec:conclusion}

We described a decentralized platform for implementing energy exchange mechanisms in a microgrid setting. Our solution enables prosumers to
trade energy without threatening their privacy or the safety
of the system. Our hybrid solver approach, which combines
a smart-contract based validator with an external optimizer,
enables the platform to clear offers securely and efficiently.

%Further, the ability to trade across multiple time intervals
%enables participants to take full advantage of batteries, thereby
%smoothening the load on the main grid. Finally, the use
%of blockchains provides decentralized trust and consensus
%capabilities, which protect from malicious actors.

In addition to the assurances provided by the distributed ledger, the resilience features provide necessary robustness for \platform. For example, $handleDeadline$ can be used to adjust $T_{lookahead}$ in order to adapt to varying complexity when there are strict timing requirements, and $handlePeerStateChange$ can be used to monitor the health of neighboring prosumers, and if they become disconnected, their trades can be removed from the smart contract.
\bibliographystyle{IEEEtran}
\bibliography{references} 

\end{document}